\documentclass{elsart3p}

\usepackage{graphicx}
\usepackage{amssymb}

\begin{document}

\begin{frontmatter}



\title{Energy Dependence of Air Fluorescence Yield measured by AIRFLY}


%\author{A. Author}
%\address{University of A, Department of B, C City}

\author[Chi]{{\bf AIRFLY Collaboration}: M. Ave},
\author[CZ1]{ M. Bohacova},
\author[LNF]{ B. Buonomo},
\author[Chi]{ N. Busca},
\author[Chi]{ L. Cazon},
\author[ANL]{ S.D. Chemerisov},
\author[ANL]{ M.E. Conde},
\author[ANL]{ R.A. Crowell},
\author[AqU]{ P. Di Carlo},
\author[RomeU]{ C. Di Giulio},
\author[CZ2]{ M. Doubrava},
\author[LNF]{ A. Esposito},
\author[Sant]{ P. Facal},
\author[ANL]{ F.J. Franchini},
\author[ANL]{J. Gebhardt},
\author[ANL]{ T. Graber},
\author[KarlsruheU]{ J. H\"orandel\thanksref{add}},
\author[CZ1]{ M. Hrabovsky},
\author[AqU]{ M. Iarlori},
\author[ANL]{ T.E. Kasprzyk},
\author[KarlsruheU]{ B. Keilhauer},
\author[FZK1]{ H. Klages},
\author[FZK2]{ M. Kleifges},
\author[ANL]{ S. Kuhlmann},
\author[LNF]{ G. Mazzitelli},
\author[ANL]{ M. Meron},
\author[CZ1]{ L. Nozka},
\author[KarlsruheU]{ A. Obermeier},
\author[CZ1]{ M. Palatka},
\author[AqU]{ S. Petrera},
\author[RomeU]{ P. Privitera},
\author[CZ1]{ J. Ridky},
\author[AqU]{ V. Rizi},
\author[RomeU]{ G. Rodriguez},
\author[AqU]{ F. Salamida\corauthref{cor1}},
\ead{francesco.salamida@aquila.infn.it}
\author[CZ1]{ P. Schovanek},
\author[ANL]{ H. Spinka},
\author[RomeU]{ E. Strazzeri},
\author[Mun]{ A. Ulrich},
\author[ANL]{ Z.M. Yusof},
\author[CZ2]{ V. Vacek},
\author[RomeS]{ P. Valente},
\author[RomeU]{ V. Verzi},
\author[ANL]{ J. Viccaro},
\author[FZK1]{ T. Waldenmaier}
% icluded to get the Institutions all in the front page
\footnotesize
\address[Chi]{ University of Chicago, Enrico Fermi Institute, 5640 S. Ellis Ave., Chicago, IL 60637, United States}
\address[CZ1]{Institute of Physics of the Academy of Sciences of the
Czech Republic, Na Slovance 2, CZ-182 21 Praha 8, Czech
Republic}
\address[LNF]{Laboratori Nazionali di Frascati dell'INFN, INFN, Sezione di Frascati, Via Enrico Fermi 40, Frascati, Rome 00044, Italy }
\address[ANL]{Argonne National Laboratory, Argonne, IL 60439 United States}
\address[AqU]{Dipartimento di Fisica dell'Universit\`{a} de l'Aquila and
INFN, Via Vetoio, I-67010 Coppito, Aquila, Italy}
\address[RomeU]{Dipartimento di Fisica dell'Universit\`{a} di Roma Tor
Vergata and Sezione INFN, Via della Ricerca Scientifica, I-00133 Roma, Italy}
\address[CZ2]{Czech Technical University, Technicka 4, 16607 Praha 6, Czech Republik}
\address[Sant]{ Departamento de F\'{\i}sica de Part\'{\i}culas, Campus Sur, Universidad, E-15782 Santiago de Compostela, Spain}
\address[KarlsruheU]{ Universit\"{a}t Karlsruhe (TH), Institut f\"{u}r Experimentelle Kernphysik (IEKP), Postfach 6980, D - 76128 Karlsruhe, Germany }
\address[FZK1]{Forschungszentrum Karlsruhe, Institut f\"{u}r Kernphysik,
Postfach 3640, D - 76021 Karlsruhe, Germany}
\address[FZK2]{Forschungszentrum Karlsruhe, Institut f\"{u}r Prozessdatenverarbeitung und Elektronik, Postfach 3640, D - 76021 Karlsruhe, Germany}
\address[Mun]{Physik Department E12, Technische Universit\"{a}t Muenchen,
James Franck Str. 1, D-85748 Garching, Germany}
\address[RomeS]{Sezione INFN di Roma 1, Ple. A. Moro 2, I-00185 Roma, Italy}
\corauth[cor1]{corresponding author}
\thanks[add]{now at Department of Astrophysics, Radboud University Nijmegen, Nijmegen, The Netherlands}

\begin{abstract}
In the fluorescence detection of ultra high energy 
($\gtrsim 10^{18}$ eV)
cosmic rays, the number of emitted fluorescence photons
is assumed to be proportional to the energy deposited in air by shower particles. We
have performed measurements of the fluorescence yield in atmospheric gases
excited by electrons over energies ranging from keV to hundreds of MeV in several accelerators. We found that within the measured energy ranges the proportionality holds at the level of few \%.
\end{abstract}

\journal{5th Fluorescence Workshop, Madrid, 2007}
\end{frontmatter}
\vspace*{1cm}
% main text
\section{Introduction}
\label{}
The detection of ultra high energy  ($\gtrsim 10^{18}$eV) cosmic rays using  
nitrogen fluorescence emission induced by extensive air showers (EAS) is a well
 established technique \cite{flyseye}. Atmospheric
 nitrogen molecules, excited by EAS charged particles (mainly $e^{\pm}$), emit
 fluorescence light in the $\approx$ 300-400 nm range.
The fluorescence detection of UHECR is based on the assumption that the number of fluorescence photons of wavelength $\lambda$ emitted at a given stage of a cosmic ray shower development, {\it i.e.} at a given altitude $h$ in the atmosphere, is proportional to the energy $E_{dep}^{shower}(h)$ deposited by the shower particles in the air volume \cite{airflyAP}:
\begin{equation}
 N_{\lambda}^{shower}(h) = E_{dep}^{shower}(h) Y_{air}(\lambda,p_0,T_0) F(\lambda,p,T),
\label{eq:nlambdash}
\end{equation}
where $ Y_{air}(\lambda,p_0,T_0)$ is the absolute yield (in number of photons per MeV) at a reference pressure $p_0$ and temperature $T_0$, $F(\lambda,p,T)$ accounts for quenching effects, and $p$ and $T$ are the air pressure and temperature at the altitude $h$. 
Since a typical cosmic ray shower extends up to about 15 km altitude, the fluorescence yield must be known over a wide range of air pressure and temperature. 
Measurements of the fluorescence yield dependence on atmospheric parameters ($F(\lambda,p,T)$) by AIRFLY are presented in a separate contribution \cite{airflyAP,paolo,paolo2}.

Simple considerations suggest that fluorescence emission should indeed be proportional to the energy deposited. In fact, the cross sections for electron excitation of the 2P and 1N nitrogen systems, which are the most relevant in the 300-400 nm range, are peaked at very low energies (tens of eV) and  decrease rapidly with energy of the electron ($\approx E^{-2}$ for the 2P and $\approx \log E/E$ for the 1N).
Therefore the fluorescence light induced by a high energy electron ($>$ keV) will be mainly produced by the secondary electrons of eV energies. 
Since the total number of secondary electrons produced by the passage of the primary electron in the air volume is roughly proportional to the energy deposited, the fluorescence light is also expected to be proportional to the energy deposited. The constant of proportionality should not depend on the primary electron energy. 

In fact, $E_{dep}^{shower}(h)$ in Eq. (\ref{eq:nlambdash}) is the sum of the energies deposited by EAS particles with a spectrum spanning from keV to GeV. 
It is thus important to verify the proportionality of the fluorescence emission to the energy deposit over a wide range of electron energies. Available measurements are limited to a few energies \cite{kakimoto} or used indirect methods \cite{flash2}.
The AIRFLY (AIR FLuorescence Yield) collaboration has performed measurements of the energy dependence of the fluorescence yield at several accelerators covering a range of electron kinetic energy from keV to hundreds of MeV.  Results of these studies are reported in the following.

\section{Electron energies from 3 to 15 MeV}
Measurements in the energy range from 3 to 15 MeV were performed at  
the Argonne Wakefield Accelerator (AWA), located at the Argonne National Laboratory. The LINAC  was operated at 5 Hz, with bunches of maximum charge of 1 nC and length 15 ps (FWHM) and typical energy spread of $\pm$ 0.3 MeV at 14 MeV. The electrons exited the accelerator vacuum through a 0.13 mm thick beryllium window. The beam spot size was typically 5 mm diameter, with negligible beam
motion. The beam intensity was monitored with an integrating
current transformer (ICT), immediately before the beam exit
flange. The signal from the ICT was integrated, digitized, and
recorded for each beam bunch. Fluorescence light produced by excitation of ambient air outside the beam exit was detected by a photomultiplier tube (Hamamatsu H7195 model) with a narrow band 337 nm filter, located about 
80 cm away from the beam axis. A shutter installed in front of
the PMT allowed measurements of background. The PMT was surrounded by
considerable lead shielding to reduce beam-related backgrounds.  The
accelerator timing signal was used to produce the integrating gate of
200 ns width. Signals were recorded using a VME standard data
acquisition system.

The LINAC was operated in a mode allowing the bunch charge to fluctuate
 over a wide range. The correlation of the PMT and ICT signals, which showed a linear relation, was fitted and the slope $S_{meas}$ was taken as
 an estimator of the fluorescence signal. The same procedure was applied with the shutter closed to estimate the background, which was subtracted . 

\begin{figure}
\begin{center}
\includegraphics [width=0.48\textwidth]{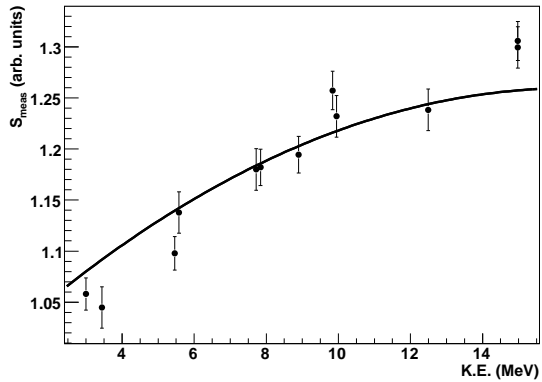}
\end{center}
\caption{Fluorescence signal as a function of kinetic energy. the full line is the result of a GEANT4 simulation where the fluorescence emission was proportional to the energy deposit.}\label{f:awarise}
\end{figure}

\begin{figure}
\begin{center}
\includegraphics [width=0.48\textwidth]{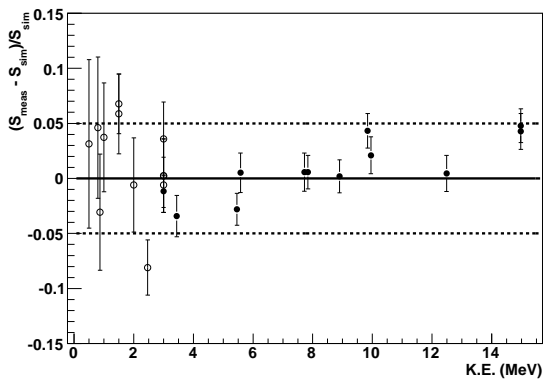}
\end{center}
\caption{Relative difference between the measured and simulated fluorescence signal as a function of kinetic energy: open dots VdG data, closed dots AWA data.}\label{f:VDG_AWA}
\end{figure}

The measured fluorescence signal $S_{meas}$ as a function of kinetic energy is shown in Fig. \ref{f:awarise}. The full line is the expected fluorescence signal, $S_{sim}$, estimated by performing a full GEANT4 simulation of the experiment.
In the simulation, the fluorescence emission was taken to be proportional to the energy deposited by the particles in the gas. Notice that the relativistic rise of the ionization losses in this energy range can be clearly seen thanks to the accuracy of our data.
The relative difference between the measured and simulated fluorescence signal,
  $(S_{meas}-S_{sim})/S_{sim}$, is shown as a function of energy in  Fig. \ref{f:VDG_AWA}. The agreement between data and the Monte Carlo simulation confirms the proportionality of the fluorescence emission to the energy deposit between 3 and 15 MeV to a level of few \%.

\section{Electron energies from 0.5 to 3 MeV}
Measurements were extended down to the minimum ionizing energy range at the Chemistry Division electron Van de Graaff (VdG) accelerator, also at the Argonne National Laboratory.
The Van de Graaff accelerator was operated in pulsed mode at 60
Hz, with beam currents from 0.2 to 0.8 $\mu$A, and nominal beam kinetic energy ranging from 0.5 MeV to 3.0 MeV. The electrons exited the accelerator vacuum through a 0.152 mm thick dura-aluminum window. The beam spot
size was typically 6 mm diameter, and a side-to-side beam motion of
approximately 5 mm was observed due to small ($<$ 1\%) variations in
the VdG energy on time scales of seconds.  Fluorescence light produced by excitation of ambient air outside the beam exit was detected by a PMT located about 60 cm away from the beam axis.
The PMT, shutter, 337 nm filter and data acquisition system were the same as in
the AWA LINAC. The beam intensity was monitored with the ICT
described before and a Faraday cup.
The total charge in the PMT was taken as a estimator of the fluorescence signal. To remove beam fluctuations,
 the PMT charge was normalized using the ICT signal. Background runs were
 also taken and substracted to the signal.

A full GEANT4 simulation of the experiment with the Van de Graaf set-up was performed, and for each energy the predicted fluorescence signal $S_{sim}$ assuming proportionality to the energy deposit was calculated.
The relative difference between the measured and simulated fluorescence signal,
  $(S_{meas}-S_{sim})/S_{sim}$, is shown as a function of energy in  Fig. \ref{f:VDG_AWA}, together with the measurements of the AWA facility. Notice that since measurements were performed at 3 MeV in both facilities, data are consistent with the proportionality of the fluorescence yield to the energy deposit with the same proportionality constant in the range 0.5 to 15 MeV. 

\section{Electron energies from 50 to 420 MeV}
\begin{figure}
\begin{center}
\includegraphics [width=0.48\textwidth]{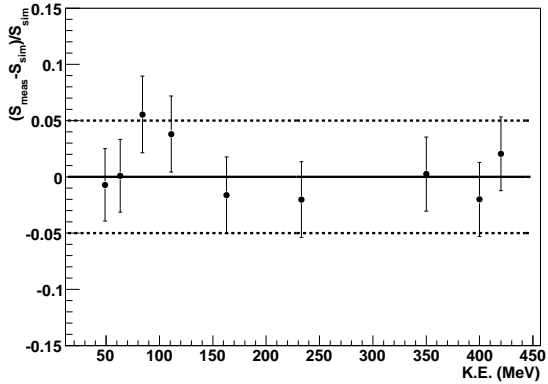}
\end{center}
\caption{Relative difference between the measured and simulated fluorescence signal as a function of kinetic energy.}\label{f:frascati}
\end{figure}

Measurements in the energy region of hundreds MeV  were performed at the BFT (Beam Test Facility) of the INFN Laboratori Nazionali di Frascati, which can deliver 50 to 800~MeV electrons and 50 to 550~MeV positrons with
intensity from single particle up to 10$^{4}$ particles per bunch at a
repetition rate of 50~Hz. The typical pulse duration was 10~ns.
The beam exited the vacuum pipe through a 0.5 mm beryllium window, and produced fluorescence light inside an aluminum pressure chamber (for a detailed description of the chamber see \cite{airflyAP}). Given the low intensity of the beam (a few $10^3$ electrons/bunch), a hybrid photodiode (HPD) with very good single photoelectron resolution was used to detect the fluorescence light. A 337 nm interference filter was placed in front of the HPD, together with a shutter that could stop the light for background measurements.  
The beam intensity was monitored by NaI(Tl) calorimeter with excellent single electron resolution, placed at the end of the beam line.
A fast scintillator was also used to monitor the beam intensity.
The dependence of fluorescence light on the primary particle energy
was measured in pure nitrogen in the range 50 to 420~MeV. We used nitrogen to increase the fluorescence light yield, given the low beam intensity. The beam
multiplicity was kept approximately constant at the individual energy
points.
The fluorescence signal  $S_{meas}$ was estimated from the number of photoelectrons measured with the HPD, after background subtraction and normalization for the beam intensity. 
The relative difference between the measured and simulated fluorescence signal,
  $(S_{meas}-S_{sim})/S_{sim}$, is shown as a function of energy in  Fig. \ref{f:frascati}, where $S_{sim}$ is the expected signal estimated by a GEANT4 simulation of the BTF set-up with the assumption of proportionality to the energy deposit. The agreement between data and the Monte Carlo simulation confirms the proportionality of the fluorescence emission to the energy deposit between 50 and 420 MeV to a level of few \%.

%%%%%%%%%%%%%%%%%%%%%%%%%%%%%%%%%%%%%%%%%%%%%%%%%%%%%%%%%%%%%%%%

\section{X-rays from 6 to 30 keV}
\begin{figure}
\begin{center}
\includegraphics [width=0.48\textwidth]{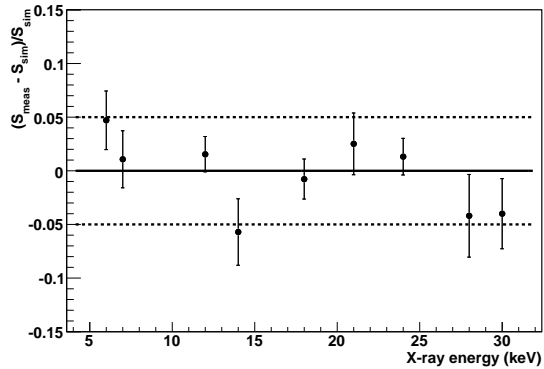}
\end{center}
\caption{Relative difference between the measured and simulated fluorescence signal as a function of X-ray energy.}\label{f:APS}
\end{figure}

Fluorescence measurements with keV electrons were performed at the Advanced Photon Source (APS) of the Argonne National Laboratory. The intense synchrotron X-ray beam of the APS 15-ID line, after exiting the vacuum beam pipe to enter the experimental hall, produced an almost monochromatic beam of electrons through photoelectric and Compton interactions with the ambient air. Electrons of energies between 6 to 30 keV produced with this method deposit all their energy in a few mm of air. The fluorescence light induced by these electrons in the ambient air was detected by the photomultiplier, 337 nm filter and shutter system previously described, placed at 9 cm distance from the beam axis. The average charge recorded by PMT was taken as an estimator of the fluorescence signal, after background subtraction. The X-ray beam intensity was monitored by ionization chambers placed along the beam axis.      
A full GEANT4 simulation of the set-up, including the ionization chambers, was performed. The relative difference between the measured and simulated fluorescence signal,
  $(S_{meas}-S_{sim})/S_{sim}$, is shown as a function of the X-ray energy in  Fig. \ref{f:APS}. Both for data and simulation, the fluorescence signal was normalized to the ionization chamber signal. There is very good agreement between data and simulation, assessing the proportionality of the fluorescence emission to the energy deposit between 6 and 30 keV to a level of few \%.

\section{Conclusions}
We presented measurements of the energy dependence of the fluorescence yield
performed at several accelerators. We tested the proportionality of the fluorescence light to the energy deposited at a level of few \% over the energy ranges 0.5 to 15 MeV, 50 to 420 MeV and 6 to 30 keV.
Notice that we performed only relative measurements within each range, and absolute measurements of the fluorescence yield are in principle needed to verify that the proportionality constant is the same in the three measured energy ranges \cite{paolo2}.  Work in this direction is ongoing. On the other hand, given that the basic mechanism for the fluorescence yield is excitation by very low energy secondary electrons, it is hard to find any physical mechanism which could change the proportionality constant between 15 and 50 MeV. The AIRFLY data presented here would then indicate that the fluorescence yield is indeed proportional to the energy deposit for electron energies at least between 0.5 and 420 MeV. Most of the EAS energy is deposited by shower particles within this energy range.

\section{Acknowledgments}
We thank the staff of Argonne National Laboratory for their support. This work was also supported by the grant of MSMT CR LC 527 and 1M06002 and ASCR grants AV0Z10100502 and AV0Z10100522. A.\ Obermeier and J.\ R.\ H\"orandel acknowledge the support of VIHKOS, which made the participation at the measurement campaigns possible.

\
% \appendix

% \section{}
% \label{}

%\begin{thebibliography}{00}


% notes:

% \bibitem{label2}
% If there is a note, it should come last:
% \bibitem{label3} \note
% \end{subbibitems}

%\bibitem{}

%\end{thebibliography}

\bibliographystyle{elsart-num}
\bibliography{biblio}

\begin{thebibliography}{00}
\bibitem{flyseye} R.M. Baltrusaitus {\it et al.}, Nucl. Instrum. Meth. Phys. Res.
   A 240 (1985) 410; T. Abu-Zayyad {\it et al.}, Nucl Instrum. Meth. Phys. Res. A 450
   (2000) 253; J. Abraham {\it et al.}, Nucl Instrum. Meth. Phys. Res. A 523
   (2004) 50
\bibitem{airflyAP}AIRFLY Collaboration, M. Ave {\it et al}, Astropart. Phys. 28 (2007) 41.
\bibitem{paolo} AIRFLY Collaboration, M. Ave {\it et al}, proceedings of this Workshop.
\bibitem{paolo2} AIRFLY Collaboration, M. Ave {\it et al}, proceedings of this Workshop.
\bibitem{martina} AIRFLY Collaboration, M. Ave {\it et al}, proceedings of this Workshop.
\bibitem{kakimoto} F. Kakimoto {\it et al.}, Nucl. Instrum. Meth. Phys. Res. A 372 (1996) 527;  P. Colin {\it et al.},  Astropart. Phys. 27 (2007) 317.
\bibitem{flash2}J. Belz {\it et al.}, Astropart. Phys. 25 (2006) 57.
\end{thebibliography}

\end{document}